\documentclass[aps,superscriptaddress,twocolumn,longbibliography]{revtex4-1}
\usepackage{amsmath}
\usepackage{amssymb}
\usepackage{graphicx}
\usepackage{subcaption} 
\usepackage[usenames,dvipsnames]{xcolor}
\usepackage{tikz}
\usepackage{pgffor}
\usepackage{verbatim}
\usepackage{float}
\usepackage{mathrsfs}
\usepackage[colorlinks, breaklinks=true,linkcolor=red, citecolor=blue, linktocpage=true]{hyperref}
\usepackage{cleveref}
\usepackage{tensor}
\usepackage{diagbox}
\usepackage[version=4]{mhchem}

\usepackage{color}

\newcommand*{\citen}[1]{%
  \begingroup
    \romannumeral-`\x 
    \setcitestyle{numbers}%
    \cite{#1}%
  \endgroup   
}


\begin{document}

\title{Sign change in the anomalous Hall effect and strong transport effects in a 2D massive Dirac metal due to spin-charge correlated disorder}


\author{Ayd\i n Cem Keser}
\affiliation{School of Physics, University of New South Wales, Kensington, NSW 2052, Australia}
\affiliation{Australian Research Council Centre of Excellence in Low-Energy Electronics Technologies,The University of New South Wales, Sydney 2052, Australia}
\author{Roberto Raimondi}
\affiliation{Dipartimento di Matematica e Fisica, Universit\`{a}  Roma Tre, Via della Vasca Navale 84, I-00146 Roma, Italy}
\author{Dimitrie Culcer}
\affiliation{School of Physics, University of New South Wales, Kensington, NSW 2052, Australia}
\affiliation{Australian Research Council Centre of Excellence in Low-Energy Electronics Technologies,The University of New South Wales, Sydney 2052, Australia}

\date{\today}

\begin{abstract}
	The anomalous Hall effect (AHE) is highly sensitive to disorder in the metallic phase. Here we show that statistical correlations between the charge-spin disorder sectors strongly affect the conductivity and the sign/magnitude of AHE. As the correlation between the charge and gauge-mass components increases, so does the AHE, achieving its universal value, and even exceed it, although the system is an impure metal.The AHE can change sign when the anti-correlations reverse the sign of the effective Dirac mass, a possible mechanism behind the sign change seen in recent experiments.	 
\end{abstract}	

\maketitle



The anomalous Hall effect (AHE) is the Hall response due to an internal magnetization, e.g. as in a ferromagnet~\cite{Nagaosa_review, QAHE_review}. It arises due to time reversal symmetry breaking  and band inversion due to strong spin orbit coupling, which on its own leads to  spin Hall effect~\cite{Bernevig}. The AHE has been predicted and observed in a variety of settings such as gapped surface modes of topological insulators~\cite{QAHE_review,TI_predict,experimental_AHE,precise_AHE, QAHE_exp1, QAHE_exp2, QAHE_exp3}. The topological effects that give rise to the quantized version of the AHE provide a paradigm for a dissipationless topological transistor~\cite{Qian1344,Liu2013,Pan2015,Vandenberghe2017,Collins2018}, which has partly motivated the great degree of interest in Dirac materials recently~\cite{shen2017topological,bernevig2013topological, dirac_materials}. 

A simple continuum model that captures many features of the quantized Hall effects is the two-dimensional (2D) Dirac Hamiltonian with the mass $m$ breaking time reversal symmetry: $H = v_F \pmb{\sigma} \cdot \pmb{p} + m \sigma_z$, where $v_F$ denotes the Fermi velocity, ${\pmb \sigma}$ the vector of Pauli spin matrices, and ${\pmb p}$ the two-dimensional momentum. 
Unlike in the insulating phase, Chern insulator~\cite{Witten, Jungwirth,Haldane,Nagaosa_japan,Haldane_model}, in the metallic phase, when the Fermi level is in the conduction or valence bands, the transport is sensitive to disorder effects~\cite{Sinitsyn, Sinitsyn2, Nunner, Bruno,Ado, Ado_long}. The disorder can manifest itself as random spatial variations in local chemical potential, Dirac mass and gauge fields~\footnote{Random fluctuations in velocity is possible due to fluctuating Rashba couplings~\cite{Dugaev,Bindel2016}, but is not considered in this letter}. The combined effect of these on the Dirac fermion  is captured by a disorder potential of the form $\tilde{\phi} \mathbf{1} + \tilde{m} \sigma_z + \tilde{\mathbf{A}}\cdot \pmb{\sigma}$, with $\tilde{\mathbf{A}}$ in the plane of the surface. Despite the long history of the AHE~\cite{Nagaosa_review}, the standard treatment focuses mostly on scalar disorder only~\footnote{See Refs.~\cite{Ado,Ado_long} for a modern treatment}. In cases where all three types are systematically treated, e.g. as in Ref.~\cite{Ludwig_IQHE}, the statistical correlations between the disorder components are not taken into account. Their complete study is a complex technical challenge. 

In this Letter, we overcome this challenge by using analytical techniques, detailed in the supplement. We first argue that even though each and every one among  $\tilde{\phi},\tilde{m},\tilde{\mathbf{A}}$ can be 
assumed short range, the components 
must be statistically correlated. We show that these correlations 
have a strong effect on transport in the metallic phase, including the sign change 
of the 	AHE, which might explain recent experiments where the sign of AHE is opposite to that of the intrinsic contribution~\cite{Checkelsky2012,Zhang1582,ChangPRL2014,LeePRB2014,Liu2018}.

\begin{table*}
	\centering
	\begin{tabular}{llllllll} 
		\hline\hline
		Disorder  is & $\hat{V}\sim$ &  $\beta$ & $\tilde{r}$                                                                                                & $\tilde{M}$ & $-2h \sigma_{xy}e^2$ & $h^2\sigma_{xx}/(2\pi e^2 \epsilon \tau)$ & $\sigma_{xx}/\sigma_{yy}$  \\ 
		\hline
		scalar (pure charge)                           & $\mathbf{1}$                             & $0$                     & 1                                                                                                                             & $m/\epsilon$  & See Eq.~\eqref{transverse}                                & See Eq.~\eqref{longitudinal}                                                                                                  & 1                                                            \\ 
		
		pure mass                                       & $\sigma_z$                                & $0$                     & $\mp 1$                                                                                                                            & $(m/\epsilon)^{\pm1}$ & See Eq.~\eqref{transverse}                               & See Eq.~\eqref{longitudinal}                                                                                                  & 1                                                            \\ 
		
		totally anti-/correlated mass-charge                & $\mathbf{1} \pm \sigma_z$ & $\pm\pi/4$  & $0$                                                                                                                             & $\pm1$                          & $\pm1$                                      & $(\epsilon \mp m)/(\epsilon \pm m)$                                              & 1                                                            \\

		mass-charge corr. critical point&  \multicolumn{3}{l}{$\beta = -\arctan(m/\epsilon)$} & $0$                          & 0                                      & $(\epsilon^2 + m^2)/(2\epsilon^2 -2m^2)$~\footnote{$r=0$ value is given, to have sign change $|r|<(\epsilon^2-m^2)/(\epsilon^2+m^2)$}             & $1$                                                           \\ 
		
		pure gauge~\footnote{Anisotropy decreases with additional gauge disorder in the $y$-direction, this does not effect $\sigma_{xy} = 0$.}                                   & $ \sigma_x$ & $0$ & $-1$	 & $m/\epsilon$                                                                                                                                                                            & $0$                                      & $1$                                                                                                        & $3$                                                            \\ 
		
		totally anti-/correlated charge-gauge & $\mathbf{1}\pm \sigma_x$&$\pm \pi/4$& $-1$&$m/\epsilon$                                                                                                                       & $\text{sgn}(m)$                                      & $(\epsilon-|m|)/|m| $               & $|\epsilon/m|$                \\
		\hline\hline
	\end{tabular}
	
	\caption{The various limits of the conductivity tensor. $\epsilon>0$ is assumed. Furthermore, one can choose $m>0$ as in the text and use the symmetries in Sec.~\ref{sec:syms} in the Supplemental material to deduce the $m<0$ and/or $\epsilon<0$ cases. }
	\label{table:limits}
\end{table*}

The 2D Dirac fermion is realized on a magnetically doped surface of a topological insulator~\cite{QAHE_review,TI_predict,experimental_AHE,precise_AHE, QAHE_exp1, QAHE_exp2, QAHE_exp3}.
The magnetic impurities, though ferromagnetically aligned, induce a random magnetization due to being randomly distributed. Furthermore, they also cause non-magnetic scattering effects by virtue of being foreign atoms in the crystal, just like other non-magnetic impurities in the system. In this way, the magnetic impurities contribute to both scalar (charge) and mass/gauge (magnetization) disorder and thus these components are correlated.  

First we concentrate on purely out-of-plane magnetization ($\tilde{\mathbf{A}}$ = 0) which leads to a charge-mass type of disorder where $\tilde{\phi}, \tilde{m}$ are zero-average short-range disorder potentials with $\langle \tilde{\phi} \tilde{m}\rangle \neq 0$. These can be parametrized in terms of their total strength, determining the mean free time $\tau$, their relative strength $r$ and correlation coefficient $s$ where $|s|,|r|<1$ . We define the correlation angle $\beta$ through $\sin(2\beta) = s\sqrt{1-r^2}$ in the region $|\beta|<\pi/4$ and  show that it has a strong impact on transport. Indeed, as we show in Sec.~\ref{app:lagrangian_absorption} of the supplement, $\beta$ can be absorbed into the effective dimensionless mass parameter $\tilde{M}$, effective relative strength $\tilde{r}$ of scalar to mass disorder and an effective mean-free time $\tilde{\tau}$, whose relations to the bare parameters are given by
\begin{equation}
\tilde{M} = \frac{m + \epsilon \tan(\beta)}{\epsilon + m\tan(\beta)}, \quad \tilde{r} = \frac{r}{\cos(2\beta)},\quad \tilde{\tau} = \frac{\tau}{\cos(2\beta)}.
\end{equation}
We assume that the Dirac mass $m$ and Fermi energy $\epsilon$ obeys $\epsilon > m > 0$ without loss of generality~\footnote{See Sec.~\ref{sec:syms} of the Supplemental material}.
Finally we compute the transport coefficients, to obtain
\begin{subequations}
	\label{analytic}
	\begin{align}
	\label{transverse}
	\sigma_{xy} &=  -\frac{4 e^2\tilde{M}}{h} \frac{[1-\tilde{r}] + [1+\tilde{r}]\tilde{M}^2}{([2-\tilde{r}] + [2+\tilde{r}]\tilde{M}^2)^2}, \\ \label{longitudinal} \sigma_{xx} &= \frac{2\pi e^2 \epsilon \tilde{\tau}} {h^2}  \frac{1-\tilde{M}^2}{2-\tilde{r} + (2+\tilde{r}) \tilde{M}^2},
	\end{align}
\end{subequations}
as though the correlations were absent, albeit included in these new parameters.

Different limits of this formula and the more general case with mass as well as gauge field disorder are summarized in Table~\ref{table:limits}. In the limit $\beta \to 0$ and $r \to 1$, the disorder becomes purely scalar, and Eq.~\eqref{analytic} reproduces the formulas previously obtained in the literature~\cite{Ado}. We note that $\sigma_{xy}$ is mostly controlled by the ratio of mass to Fermi energy $m/\epsilon$. For example in the scalar disorder limit, $\sigma_{xy}$ decays with $(m/\epsilon)^3$ as $m/\epsilon \to 0$, and monotonically approaches to the universal value $-e^2/(2h)$ as $m \to \epsilon$ as expected. In the case of finite correlations, the same behavior applies but this time controlled by the effective mass parameter $\tilde{M}$. This means that the disorder correlations can change the sign and magnitude of the transverse conductivity to a great extent. For example, positive correlations, $\beta >0$ increase $\tilde{M}$ hence the magnitude of $\sigma_{xy}$. Perfect correlations require that the intrinsic non-magnetic sources of disorder vanish, which is physically unrealistic. In this limit, however, we have $r = 0,\: \beta \to \pi/4$ which pushes the effective mass $\tilde{M}$ to $1$ so that $\sigma_{xy}$ hits the universal AHE value despite the system being an impure metal. On the other hand, when the mass and charge components are anti-correlated, $\beta < 0$, and the effective mass $\tilde{M}$ together with $\sigma_{xy}$ switches sign for $\beta < -\arctan(m/\epsilon)$. Finally, when the mass-charge components are completely anti-correlated, $ r=0, \:\beta \to -\pi/4$ and $\tilde{M}\to -1$, and hence regardless of the sign of mass $m$, the transverse conductivity achieves its half-quantized value with the opposite sign $e^2/(2h)$. Since charge-mass correlations break time reversal symmetry, the AHE remains finite as $m/\epsilon \to 0$ and has the same sign as $\beta$. When the disorder is an uncorrelated combination of mass and charge, we have  $\beta= 0$, and this signature disappears along with the possibility of sign change. 

Correlations have a profound effect on the longitudinal transport as well. In the absence of correlations, $\beta = 0$, the longitudinal conductivity is proportional to the mean free time and increases with $\epsilon/m$ as expected. Correlations exert their influence on both of these parameters. For instance, if it was not that the effective mean free time diverges, maximum correlations $r=0,\: \beta \to \pi/4$ would lead effectively to a Fermi level at the band edge situation, $\tilde{M} = 1$ which would make the conductivity zero. However, evaluating the  limiting value of the product $\tilde{\tau}(1-\tilde{M}^2	)$, we find that $\sigma_{xx} \sim \tau (\epsilon-m)/(\epsilon + m)$, hence must vanish when the bare Fermi level hits the band edge. On the other hand, the $\epsilon \to m $ limit is singular in the presence of anti-correlations. Taking this limit at the out-set gives $\sigma_{xx} = 0$ however, for the value $\beta = -\arctan(m/\epsilon)$, where the anti-correlations nullify the effective mass, assuming $r = 0$ for simplicity, $\sigma_{xx} \sim \tau (\epsilon^2+m^2)/(\epsilon^2 - m^2)$, which diverges in the gapped limit. Similarly, when the charge-mass components are completely anti-correlated we get $\sigma_{xx} \sim \tau (\epsilon+m)/(\epsilon - m)$. This signals that, in case of anti-correlations the longitudinal conductivity increases as the Fermi level decreases towards the mass gap, up to the point where the good metal approximation, within which we confine ourselves here, breaks down. Around the metal-insulator transition disorder leads to formation of bounds states and a percolated domain structure~\cite{Ludwig_IQHE} and the effect of correlations on this will be studied in a future publication.   

So far Eq.~\eqref{analytic} only captures correlated mass-charge disorder. In general, even when the gauge field component of disorder $\tilde{\mathbf{A}}$ is finite and correlated with mass and charge components, we find, by direct evaluation of diagrams as described in Sec.~\ref{sec:diagrams} of the supplement, that stronger correlations between scalar and non-scalar components increase the AHE, and in the limit of perfectly correlated and equal strength charge-gauge $\mathbf{1}\pm \sigma_x$ field disorder, the AHE, to leading order $\tau^0$, assumes its universal value $-\text{sgn}(m)e^2/(2h)$ once again, despite the system being an impure metal. Anti-correlations have the same effect in this type of disorder, as the sign of correlations coefficient can be inverted with a rotation. Therefore such disorder cannot change the sign of the  AHE which always assumes the sign of bare average magnetization  i. e. mass $m$.  The subtle effect of correlations increases the magnitude of the Hall conductivity even beyond the universal value up to $\approx 0.7 e^2/h$, for example when the disorder is $\sim \sqrt{3} \mathbf{1}/2 + \sigma_x/2$ and $m/\epsilon \approx 0.4$. 

Notably, the existence of a random gauge field that breaks rotational symmetry has the anisotropic  transport signature $\sigma_{xx}\neq\sigma_{yy}$. When the disorder is of purely gauge form, say $\sim \sigma_x$, we find $\sigma_{xx}/\sigma_{yy} = 3$ and in the case of equal strength and completely  correlated components, $\sim \mathbf{1}+\sigma_x$ we have $\sigma_{xx}\sim \tau(\epsilon-m)/m$ and $\sigma_{xx}/\sigma_{yy} = \epsilon/m$. Therefore such correlations would manifest as enhanced longitudinal conductivity in the alignment direction, especially in the deep metallic regime $\epsilon \gg m$.

When the random gauge field is uncorrelated with the mass-charge components, we can reduce the disorder to a combination of uncorrelated components. First we eliminate the mass-charge correlations by absorbing them into the parameters as in Eq.~\eqref{analytic}. Then we can set $\langle \tilde{A}_x \tilde{A}_y\rangle = 0$, without loss of generality, by a suitable choice of coordinates and exploiting the gauge invariance~\footnote{See Sec.~\ref{app:gauge_correlation} of the supplemental material for details}. An uncorrelated random gauge field suppresses the AHE by switching the sign of $m$ in the electron disorder self-energy. Ultimately, in the limit of pure gauge disorder $\hat{V} \sim \sigma_x$, the AHE vanishes.

\begin{figure}[h]	
	\centering
	\includegraphics[width=0.4\textwidth]{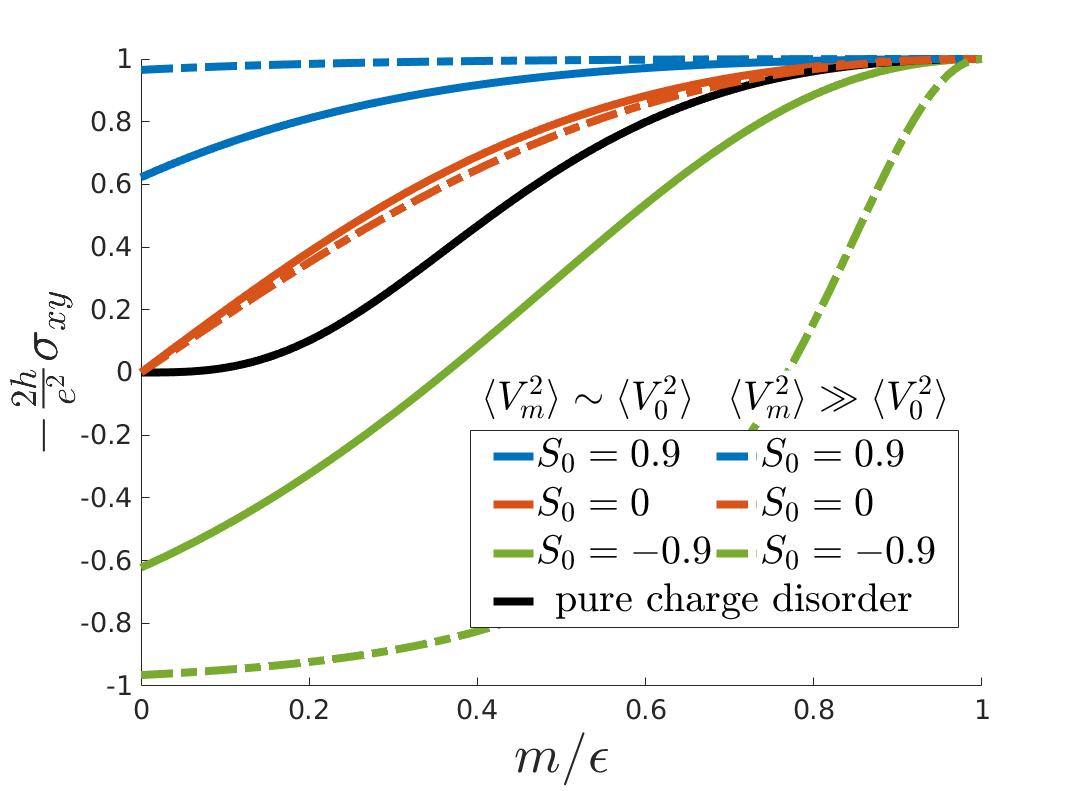}
	\caption {The anomalous Hall conductivity divided by its universal value, for various values of the coefficient $S_0$  that measures the scalar component of magnetic impurities in Eq~\eqref{disorder_physical}.The dashed lines depict the AHE when the magnetic dopant concentration, so the strength of the associated disorder term $\langle V_m^2\rangle$ is very high compared to  strength of non-magnetic impurities $\langle V_0^2\rangle$. }
	\label{fig:s_xy_vs_m}
\end{figure}

\begin{figure}[h]	
	\centering
	\includegraphics[width=0.5\textwidth]{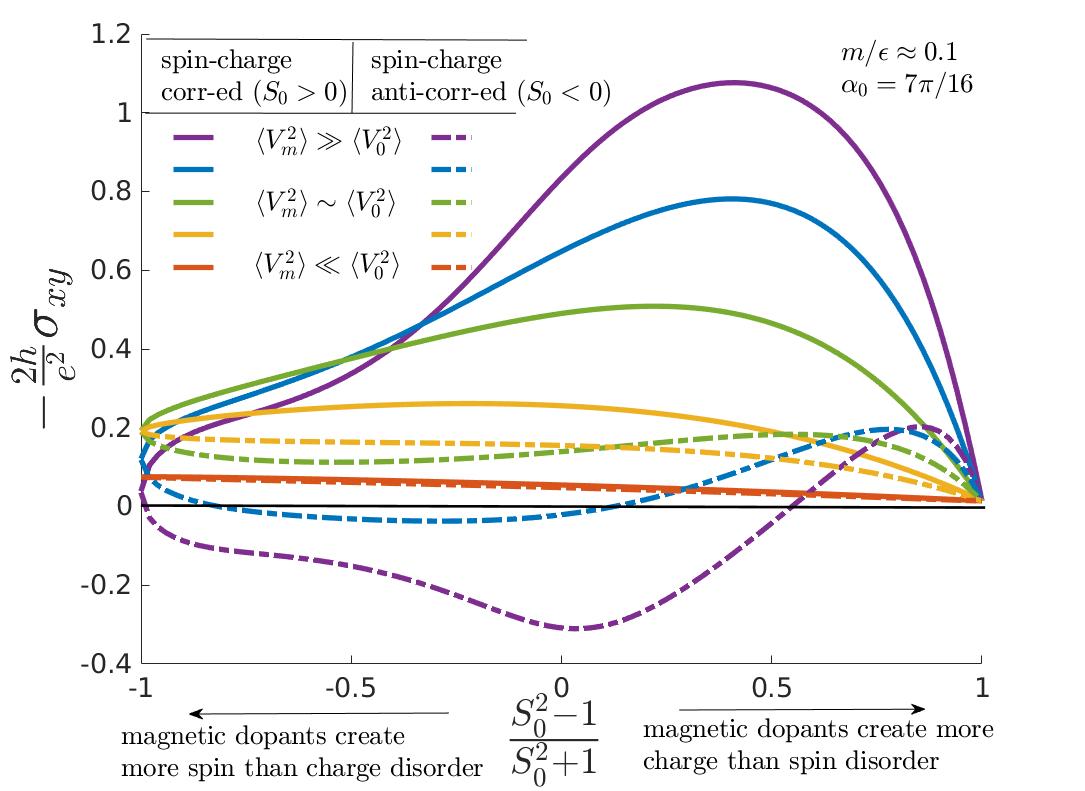}
	\caption {The anomalous Hall conductivity divided by its universal value versus the relative strength of charge to mass/gauge disorder created by magnetic impurities, for various values of the relative strength of magnetic to non-magnetic impurities. Solid and dashed lines are when magnetic impurities create repulsive and attractive ($S_0>0$, $S_0<0$) charge disorder respectively. See Eq.~\ref{disorder_physical}.  }
	\label{fig:s_xy_vs_r1}
\end{figure}

We argue that such disorder comes about naturally if the system is realized on the surface of a 3D topological insulator with magnetic impurities~\cite{massive_Dirac}. The impurity spins $\vec{m}$ tend to align ferromagnetically due to interactions mediated by surface Dirac fermions through the RKKY mechanism (when there are free carriers) and through the Van Vleck mechanism in the insulating phase~\cite{TI_predict,MingdaPRL2015}. This vector picks the easy axis of magnetization that makes an angle $\alpha_0$ with the surface normal. This axis  is generally out-of-plane on the surface~\cite{ferro_perp_Hasan,ferro_inplane, ferro_perp} and in-plane in the bulk, but we solve the case of arbitrary $\alpha_0$, so as to allow for thin film effects. The average out-of-plane magnetization $\langle m_z \rangle $ is the mass gap (simply $m$, and assumed $m>0$), while the term $\tilde{m} = m_z - \langle m_z \rangle$ is a zero-average disorder term that shares the same statistics with the distribution of magnetic impurities. Since this random magnetization appears as a random Dirac mass in the Hamiltonian, we will refer to it as mass disorder. The mean value of the in-plane component of the dopant spins can be gauged away, and thus one is left with random $ \tilde{A}_x = m_x - \langle m_x \rangle $ and $\tilde{A}_y = m_y - \langle m_y \rangle$ terms that couple to the fermion like a gauge field would. Finally, the magnetic impurities, being foreign atoms, disturb the underlying lattice thereby creating a random electric potential. The mean value of this is counted in the Fermi level, the remainder contributing to a zero-average random electric potential $\tilde{\phi}$. 
In addition, the scalar component of disorder $\tilde{\phi}$, must contain the effects typically due to imperfections, dislocations and defects that we call $V_0$. All in all, the disorder potential can be written as 
\begin{equation}
\label{disorder_physical}
\hat{V} = V_0\hat{1} + V_m (S_0 \hat{1}+ \cos(\alpha_0) \sigma_z + \sin(\alpha_0) \sigma_x ).
\end{equation}
Here, $V_0,V_m$ are zero mean uncorrelated short-range disorder that satisfies $\langle V_{0,m}(\mathbf{x})V_{0,m}(\mathbf{x}')\rangle = (1\pm R_0)\delta(\mathbf{x}-\mathbf{x}')/\tau$ respectively, so that their relative strength is $R_0$. $S_0$ parametrizes the strength of scalar disorder created by the magnetic atoms and can in principle take any value. Depending on the location and type of these atoms, $S_0$ has either sign. For transition metal doped TI's, the metal atoms create attractive centers for electrons, thereby creating charge-mass anti-correlated disorder that is , $S_0,\beta<0$~\cite{QAHE_exp2}.  As long as the spins of all the impurities are locked in, we can put the $\sigma_y$ component of disorder to zero without loss of generality. The random electric field, mass and gauge fields arising due to this potential are $\tilde{A}_x = V_m \sin(\alpha),
\tilde{m} = V_m \cos(\alpha)$ and $
\tilde{\phi} = V_0 + V_m S_0 $ respectively. Therefore, we find out that, even if we assumed the magnetic and intrinsic disorders, $V_m$ and $V_0$ were uncorrelated, mass, gauge and charge sectors are necessarily correlated, as they  have the random variable $V_m$ in common.

If the magnetization is perfectly out-of-plane (and when the effect of possible in-plane-magnetization in the 3D, TI bulk is ignored), the gauge disorder vanishes. In this case, the relative strength of charge $\tilde{\phi}$ to mass disorder $\tilde{m}$ is extracted as $r = (2R_0 + [1-R_0]S_0^2)/(2+[1-R_0]S_0^2)$ and the correlation angle $\beta$ satisfies $\sin(2\beta) = 2S_0(1-R_0)/(2+[1-	R_0]S_0^2)$. The conductivity is then given by the closed form expressions in Eq.~\eqref{analytic}. Correlations/anti-correlations increase as $S_0 \sim \pm 1$ respectively, and when the magnetic dopant concentration rises or $R_0\to -1$. Also, the AHE changes sign whenever $m + \epsilon\tan(\beta) = 0$.

As seen in Fig.~\ref{fig:s_xy_vs_m}, when the magnetic dopants create no charge disorder, $S_0 = 0$, the resulting mass/gauge disorder is uncorrelated with the charge component and affects the	 magnitude of the AHE slightly, as compared to the pure charge disorder case. However, once the charge component of magnetic dopants is taken into account, correlations/anti-correlations between charge and mass components emerge. Positive correlations improve the AHE significantly, while anti-correlations reverse its sign. As the dopant concentration rises, so does the strength of correlations and their effect on the AHE. The tilting angle suppresses these effects as it decreases the mass component of disorder hence its correlations/anti-correlations with the charge disorder.  

The correlation effect is large enough even to compensate the decay of total magnetization. For example,as seen in Fig.~\ref{fig:s_xy_vs_r1} when the mass gap is given by the average out-of-plane magnetization of the impurity spins, thereby obeying $m \sim \cos(\alpha_0)$, the AHE becomes insensitive to a wide range of tilting angles $\alpha_0$, all the way up to the point where the spins are in-plane ($\alpha = \pi/2$) given that the correlations are large and the gauge/mass and scalar components have comparable strength.




Since we have studied the system in the good metal regime, our results hold well when the Fermi energy is sufficiently above the gap, but below the point where bulk conduction bands are excited. The opposite limit, where Fermi energy approaches the gap, has further complications. The screening of disorder potential gets weaker and to capture the smooth nature of disorder, non-local correlations should be taken into account, as done in Ref.~\cite{Ado_long} for charge-disorder. Moreover, the spin texture gets weaker due to non-magnetic gap effects~\cite{Liu2018} and therefore the Dirac model is not sufficient. 

We propose the disorder correlations as a possible mechanism leading to a change in the sign of AHE in the metallic regime, without changing the Dirac mass e.g. magnetization and/or exchange coupling in magnetic topological insulators. According to recent experiments, the sign of AHE can be the same~\cite{Zhang1582,Zhang2014,Chang_adv_mat} as or differ~\cite{Checkelsky2012,Zhang1582,ChangPRL2014,LeePRB2014,Liu2018} from that of the intrinsic contribution~\cite{QAHE_exp1,QAHE_exp2,QAHE_exp3,experimental_AHE}, depending on the magnetic doping concentration~\footnote{See Table~\ref{tabel:sign_problem} in the supplement for details} . An interplay of the tilting of magnetization, the position of Fermi level and the dopant concentration can explain the change of sign in the AHE observed in magnetically doped topological insulator thin films. If the Dirac mass is positive, and assuming that the magnetic dopants create attractive centers for the carriers, the mass and charge components of the disorder become anti-correlated.  When the Fermi level, the magnetization gap, intrinsic disorder and the tilting angle are fixed, if the concentration of magnetic dopants increase, so do the anti-correlations, which cause a sign change by reversing the sign of the effective mass parameter. The ionic character of the dopants might lead to stronger charge-mass correlations by increasing the parameter $|S_0|$ in our model. Therefore elements with such property might induce a sign change at lower doping concentrations. 



In summary, we have systematically studied the correlations between  charge, mass and gauge components of disorder in the 2D massive Dirac system and shown that transport and AHE in particular is very sensitive to them, despite being independent of total strength of disorder. We have shown that the correlations between mass and charge component can be absorbed into the effective mass-to-Fermi energy ratio, that controls both the sign and magnitude of the AHE. Therefore strong correlations can enhance AHE while anti-correlations induce a sign change. In the limit of perfect correlations, AHE can assume its universal value, with a sign dictated by the correlation coefficient.  Moreover, we have shown that gauge disorder, if correlated with the charge component, can significantly change the magnitude of the Hall conductivity and increase it up to the universal value and even beyond. Furthermore, gauge disorder creates an anisotropic longitudinal transport signature. However, this kind of disorder cannot, by itself, induce a sign change due to symmetry considerations. Hence the only possibility to reverse the sign of AHE without flipping the Dirac mass is to include strong correlations between mass and charge components. We have argued that such disorder comes about naturally in the experimental realization of the Dirac fermion on magnetically doped TI surfaces. The out-of-plane component of magnetization due to randomly distributed magnetic dopants create correlated mass-charge disorder, while the in-plane component creates an additional correlated gauge disorder.  We propose the correlations between disorder components as a possible mechanism to explain the `sign problem' observed in recent transport experiments on magnetically doped topological insulator thin films.

\acknowledgments

A.C.K. is grateful to Oleg Sushkov, Ali Yazdani, Yongqing Li and Yaroslav Kharkov for valuable discussions. This research was supported by the Australian Research Council Centre of Excellence in Future Low-Energy Electronics Technologies (project CE170100039) and funded by the Australian Government.


\bibliographystyle{apsrev4-1}
\bibliography{my-refsv}


\clearpage
\onecolumngrid
\setcounter{page}{1}

\part*{\centering Supplemental Material}
\setcounter{section}{0}
\section{Kubo formula and leading order diagrams}
To first order in the external electric field, the conductivity tensor can be computed by using expressions that involve the retarded and advanced Green's function of the electron system. The dominant contribution in the metallic phase comes  from the Fermi surface is expressed by the Kubo term
\begin{equation}
\sigma_{ij} =  e^2\text{Tr}\langle \hat{v}_{i} G^R \hat{v}_j G^A\rangle/h,	
\end{equation}
where $e$ is the electron charge and $h$ is the Planck's constant and $\hat{v},\:i =1,2$ are the velocity operators.  In the 2D massive  Dirac Hamiltonian $\hat{v}_i = v_F \sigma_{i}$ and the trace is carried out both 
in momentum and spinor space.

Through out the text we use units where $v_F = \hbar = |e| = 1$. 

\begin{figure}[h]
	\centering
	\begin{subfigure}{0.4\textwidth}
		\centering
		\includegraphics[width=1\textwidth]{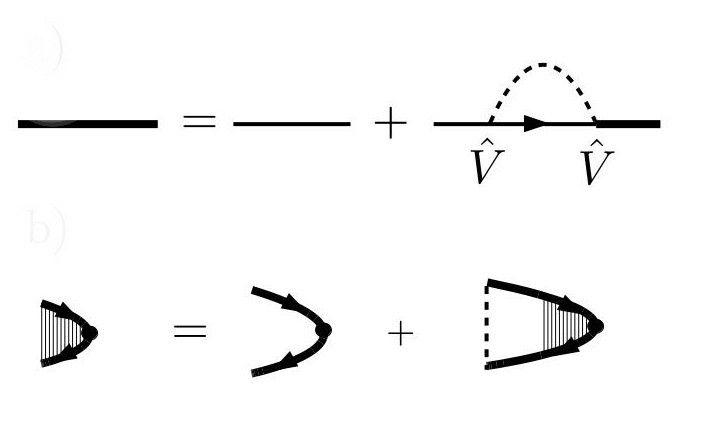}
		\caption {}
		\label{fig:self-energy-vertex}
	\end{subfigure}
	\begin{subfigure}{0.5\textwidth}
		\centering
		\includegraphics[width=1\textwidth]{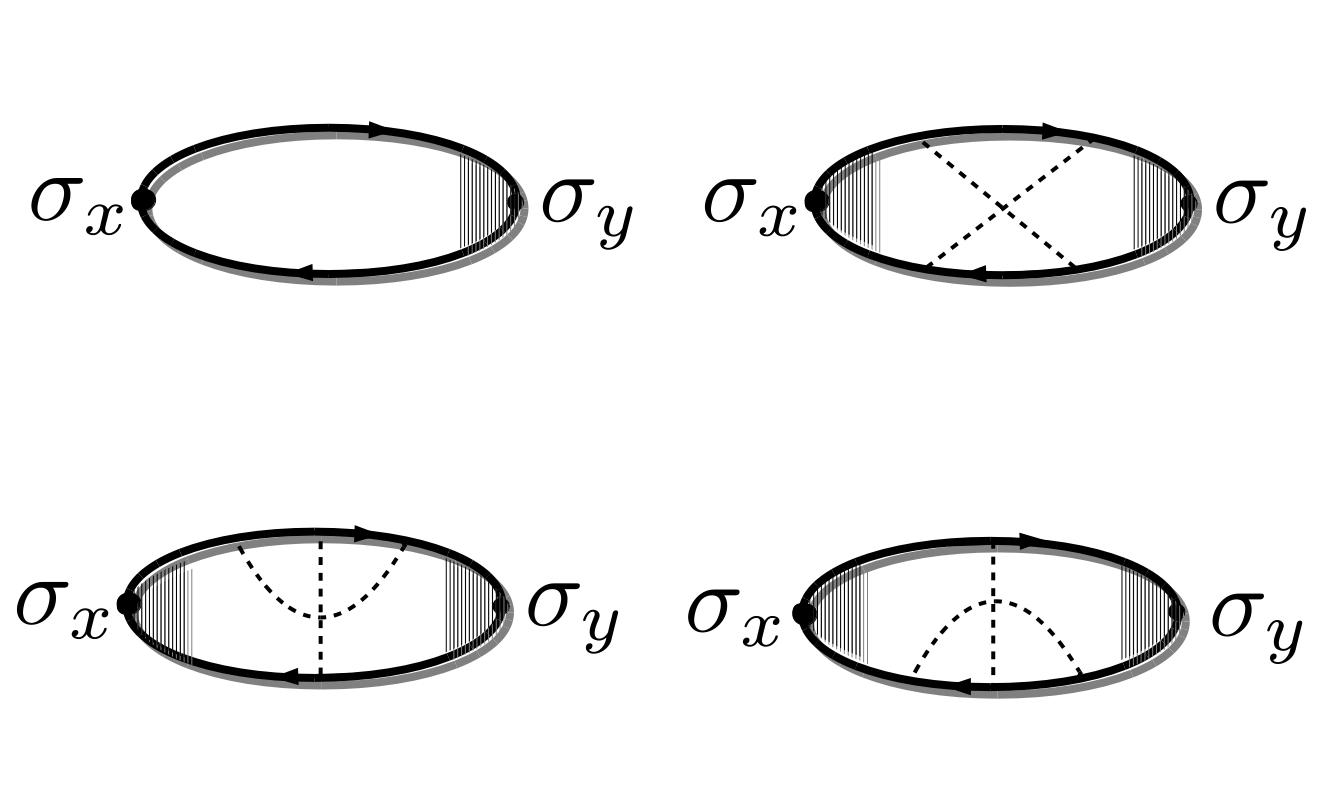}
		\caption{}
		\label{fig:bubbles}
	\end{subfigure}
	\caption{Diagrammatic summary of our calculation. a) Disorder corrections to Green's function at Born level and the non-crossed vertex correction. b) Leading order $\tau^0$ diagrams for the Hall conductivity, the non-crossed, $X$, and $\Psi$ diagrams~\cite{Ado}.  For the longitudinal conductivity only the top left non-crossed diagram contributes. }
	\label{fig:diagrams}
\end{figure}

Kubo formula can be represented by a bubble diagram where $\sigma_{i,j}$ are the vertices and the Green's functions are lines that connect them. When there is disorder, this diagram is decorated by disorder lines that carry momenta. A diagrammatic summary of our calculation is given in Fig.~\ref{fig:diagrams}.

\section{Discrete Symmetries}
\label{sec:syms}
When the sign of Fermi energy is reversed, the Green's functions obey
\begin{equation}
G^R(-|\epsilon|,m,\mathbf{k}) = (-|\epsilon|-\pmb{\sigma}\cdot \mathbf{k} - m\sigma_z + i0^+)^{-1} = -(|\epsilon|+\pmb{\sigma}\cdot \pmb{k} + m\sigma_z + i0^+)^{-1} = -G^A(|\epsilon|,-m,-\mathbf{k}).
\end{equation}
Therefore, from the Kubo formula we immediately get
\begin{equation}
\sigma_{ij}(-|\epsilon|,m) = \sigma_{ji}(|\epsilon|,-m).
\end{equation}

Moreover, the Hall conductivity is symmetric as long as all the TR breaking terms $\mathbf{V}$ in the disorder $\hat{V} = V_0 \mathbf{1} + \mathbf{V}\cdot \pmb{\sigma}$ are reversed. To see this we rotate the coordinate system by $\pi$ about the $y$-axis, i.e. $\sigma_x,\sigma_z \to -\sigma_x, -\sigma_z$ and conjugate the Kubo formula. In effect, the mass and all the correlated magnetic disorder components are reversed;
\begin{equation}
\sigma_{ij}(m,\mathbf{V}) = \sigma_{ji}(-m,-\mathbf{V}).
\end{equation}

If the system is isotropic, doing $\sigma_x, \sigma_y \to -\sigma_y, \sigma_x$ and interchanging the $x \leftrightarrow y$, shows that the Hall conductivity is anti-symmetric. Also, if all the disorder vertices can be made real (by using gauge transformations), one can conjugate the Kubo formula, and find out that 
\begin{equation}
\sigma_{xy} = -\sigma_{yx}.
\end{equation}

\section{Disorder with correlated gauge components}
\label{app:gauge_correlation}
If the random gauge field components have $\langle \tilde{A}_x \tilde{A}_y\rangle \neq 0$, the resulting out of plane magnetic field has
\begin{equation}
\label{magnetic}
\langle \tilde{B}(\pmb{k}) \tilde{B}(\pmb{k'})\rangle = -(2\pi)^2 \delta(\pmb{k}+\pmb{k'})\\ \left(k_x^2\langle \tilde{A}_y^2\rangle + k_y^2\langle \tilde{A}_x^2\rangle - 2k_x k_y \langle \tilde{A}_x\tilde{A}_y\rangle\right),
\end{equation}
where $\langle \tilde{A}_{x,y}^2\rangle = (1\pm r)/\tau$ and $\langle A_x A_y\rangle = s\sqrt{1-r^2}$. Defining $\cos(2\alpha) = r$, the expression in paranthesis in Eq.~\eqref{magnetic} reads
\begin{equation}
\frac{2}{\tau} (k_x^2 \cos^2(\alpha) + k_y^2 \sin^2(\alpha) - 2 s k_x k_y \cos(\alpha)\sin(\alpha)).
\end{equation}
Making a spatial rotation by the angle
\begin{equation}
\theta = \frac{1}{2} \arctan(s \tan(2\alpha)),
\end{equation}
and defining
\begin{equation}
\tilde{r} = r\cos(2\theta)
\end{equation}
eliminates the correlations. Therefore one can always pick a coordinate system where the components of the disordered gauge field is uncorrelated. 

\section{Arbitrarily correlated charge-mass disorder}
\label{app:lagrangian_absorption}
Suppose we have a disorder potential of the form
\begin{equation}
\hat{V} = V_0 + V_1 \hat{\sigma}_z,
\end{equation}
where $V_0$ and $V_1$ are correlated. If $V = (V_0, V_1)^T $, then the covariance matrix is
\begin{equation}
\langle V(\mathbf{x}) V^T(\mathbf{x'}) \rangle =\frac{1}{\tau}
\begin{pmatrix}
(1+r)  & s \sqrt{1-r^2} \\
s \sqrt{1-r^2} & (1-r)
\end{pmatrix}
\delta(\mathbf{x}-\mathbf{x'}),
\end{equation}
where $-1<s<1$ is the correlation coefficient, and $-1<r<1$ is the relative strength.
Since we assume all disorder potentials are correlated, we will suppress the spatial dependence of correlation functions. 
If we define
\begin{equation}
\beta = \frac{1}{2}\arcsin(s \sqrt{1-r^2}),\quad \tilde{r} = \frac{r}{\cos(2\beta)},
\end{equation}
and two uncorrelated Gaussian random variables
\begin{equation}
\langle  W^2_\pm \rangle = \frac{1}{\tau}(1 \pm \tilde{r}) ,\quad \langle W_+ W_-\rangle = 0;
\end{equation}
then the disorder potential becomes
\begin{equation}
\hat{V} = W_+ (\cos(\beta) + \sin(\beta)\sigma_z) + W_- (\sin(\beta) + \cos(\beta)\sigma_z).
\end{equation}
Assuming $\cos(2\beta)>0$ with out loss of generality, we can define the transformation
\begin{equation}
\bar{\psi} \to \bar{\psi} \Gamma^{-1/2}, \quad \psi \to \Gamma^{-1/2}\psi,
\end{equation}
where
\begin{equation}
\hat{\Gamma}  = \frac{1}{\sqrt{\cos(2\beta)}} [\cos(\beta)+\sin(\beta)\sigma_z];
\end{equation}
we get 
\begin{equation}
L \to \frac{\epsilon \cos(\beta) + m\sin(\beta)}{\sqrt{\cos(2\beta)}} - \pmb{\sigma} \cdot \mathbf{k} - \frac{m \cos(\beta) + \epsilon \sin(\beta)}{\sqrt{\cos(2\beta)}}\sigma_z  - W_+\sqrt{\cos(2\beta)} - W_- \sqrt{\cos(2\beta)} \sigma_z,
\end{equation}
and this can be rewritten as
\begin{equation}
L \to \tilde{\epsilon} - \pmb{\sigma}\cdot \mathbf{k} -\tilde{m}\sigma_z - \tilde{W}_+ - \tilde{W}_- \sigma_z,
\end{equation}
where 
\begin{equation}
\langle  \tilde{W}^2_\pm \rangle = \frac{1}{\tilde{\tau}}(1 \pm \tilde{r}) ,\quad \langle \tilde{W}_+ \tilde{W}_-\rangle = 0 ,\quad \tilde{\tau} = \tau/\cos(2\beta).
\end{equation} 	

\section{Diagrammatic calculation of conductivity with correlated short-range disorder}
\label{sec:diagrams}

In the clean metal case, the transverse conductivity is mostly accounted for by the Kubo term $\sigma_{xy}^{I}$. Diagrammatically this is represented by the bare bubble, with vertices $\sigma_x$ and $\sigma_y$. A straight forward computation gives the intrinsic contribution to Hall conductivity
\begin{equation}
\sigma_{xy}^{intr.} = -\frac{e^2}{2h} \frac{m}{|\epsilon|}.
\end{equation}

The short-range disorder $\hat{V}$, creates scattering processes. A sequence of Born scatterings is captured by the insertion of the self-energy diagram into the Green's function line as in Fig.~\ref{fig:self-energy-vertex}. Moreover, the disorder decorates the bubble diagram to yield the leading order diagrams depicted in Fig.~\ref{fig:bubbles}.

Since we can eliminate the correlations between the charge and mass components, we can transform the disorder potential in the canonical form 
\begin{equation}
\label{disorder_transformed}
\hat{V}={W}_+ (\cos({\gamma}) + \sin({\gamma})\sigma_x) + {W}_- (\cos({\alpha}) \sigma_z+ \sin({\alpha})\sigma_x),
\end{equation}
where $W_\pm$ are uncorrelated Gaussian random variables with strengths $\langle W_\pm^2\rangle = (1\pm r)/\tau$
as we show in Sec~\ref{app:lagrangian_absorption} of this supplement. We call the angles $\alpha,\gamma$ the effective angle of tilting and effective angle of correlation respectively.  Now, given this disorder model,  we calculate these diagrams one by one.


\subsection{Self-energy}
\label{sec:self-energy}
The linear dispersion Dirac model has a UV divergent real self-energy and an RG procedure renormalizes the mass, Fermi energy and the disorder strength. We discuss the details of this procedure for spin-charge correlated $\hat{V}$ in Sec~\ref{app:RG} of this supplement.

Working with the transformed potential Eq.~\eqref{disorder_transformed}  the remaining imaginary part of self energy can be written as
\begin{equation}
\hat{\Sigma}=-\frac{i}{2\tau} \left( \epsilon + b m\sigma_z +   c  \sigma_x\right),
\end{equation}
where
\begin{subequations}
	\begin{align}
	b &=  \frac{1+r}{2} \cos(2\gamma) + \frac{1-r}{2}\cos(2\alpha),\\
	c &= \frac{1+r}{2} {\epsilon} \sin(2\gamma) + \frac{1-r}{2} m \sin(2\alpha) .
	\end{align}
\end{subequations}
With this self energy disordered Green's function is
\begin{equation}
\!\!\!\!\!\!\!\!\!\!G^R = \frac{\epsilon \left(1+  \frac{i}{2\tau} \right) + \sigma_x \left(k_x - c \frac{i}{2\tau}\right) +\sigma_y k_y + m\sigma_z \left(1-b \frac{i}{2\tau} \right) }{\epsilon^2 -m^2 -k^2 +\frac{i}{\tau}[  \epsilon^2 + b m^2 + ck_x]}.
\end{equation}

The self-energy and vertex diagrams give insights as to why the limits in Table~\ref{table:limits}  of the paper come about. For example when the disorder is pure gauge, the disorder vertex flips the sign of the mass term both in the self-energy and the corrected vertices. As we will see, this makes the transverse conductivity vanish. When the disorder is $~ \mathbf{1}+\sigma_z$, the disorder vertex can be moved across a $\sigma$ vertex, that is $(\mathbf{1}+\sigma_z)\sigma_x (\mathbf{1}+\sigma_z) = 0$ and thus the corrections vanish. However, in the self-energy, moving the disorder vertex from one end to the other has no effect. Therefore, we the disorder decorated diagrams are easily calculated to yield the universal value.  

\subsection{1-Loop RG in Disordered 2D Massive Dirac Metal}
\label{app:RG}
Starting from
\begin{equation}
\hat{V} = W_1 \hat{\Gamma}_1 + W_2 \hat{\Gamma}_2,
\end{equation}
where
$
\hat{\Gamma}_{1} = A_1 + B_1 \sigma_x,\quad \hat{\Gamma}_{2} = A_2\sigma_z + B_2 \sigma_x
$
and $W_\pm$ are uncorrelated random variables with
$
\langle W_1^2 \rangle = \langle W_2^2 \rangle =  2\pi
$
The Born self energy is then
\begin{equation}
\hat{\Sigma} =  \sum_{a=1,2}2\pi \int \frac{d^2k}{(2\pi)^2} \: \hat{\Gamma}_{a} \hat{G} \hat{\Gamma}_{a};
\end{equation}
and the divergent real part reads
\begin{equation}
\hat{\Sigma}_R = -\sum_{i} \left[ \epsilon(A_i^2 + B_i^2) + m(A_i^2-B_i^2)\sigma_z + (2A_1B_1 \epsilon + 2A_2B_2m)\sigma_x \right] \int^{\Lambda} \frac{d\xi}{\xi}.
\end{equation}
The $\sigma_x$ term can be absorbed into a gauge. Mass and Fermi energy are renormalized according to

\begin{equation}
\frac{d}{d\ln \Lambda}\
\begin{pmatrix}
\epsilon \\ m
\end{pmatrix}
= 
\begin{pmatrix}
-A_1^2 - A_2^2 -B_1^2 - B_2^2 &&0 \\
0 && A_1^2 + A_2^2-B_1^2 - B_2^2	
\end{pmatrix}
\begin{pmatrix}
\epsilon \\ m
\end{pmatrix}.
\end{equation}


\begin{figure}[h]
	\centering
	\includegraphics[width=0.4\textwidth]{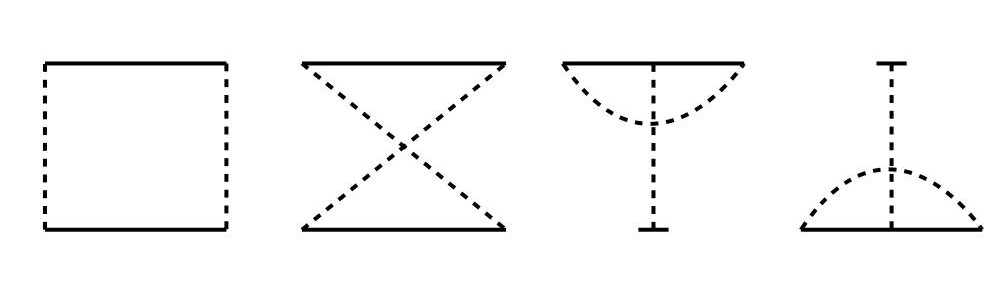}
	\caption {The UV divergent 1-loop diagrams that renormalize the disorder vertex strengths and eliminate the UV divergent Born self-energy}
	\label{fig:RG}
\end{figure}

The divergent contributions from the diagrams in Fig.~\ref{fig:RG} renormalize the vertex strengths as
\begin{subequations}
	\label{vertex_renorm}
	\begin{align}
	-\frac{d \ln A_1}{d\ln \Lambda} &= A_1^2 + B_1^2 +A_2^2 + B_2^2, \\
	-\frac{d \ln B_1}{d\ln \Lambda} &= 2A_1^2, \\	
	\frac{d \ln A_2}{d\ln \Lambda} &= A_1^2 - B_1^2 +A_2^2 - B_2^2, \\
	\frac{d \ln B_2}{d\ln \Lambda} &= 2A_2^2.	
	\end{align}
\end{subequations}
As long as the interaction strength is $A_0^2 \ln\Lambda_0, B^2\ln\Lambda_0 \ll 1$, these equations can be solved approximately, and the UV divergence is eliminated in return for a slight change in system parameters. When the $x$-component of the disorder is killed, these equations take the expected form.
Writing 
\begin{subequations}
	\begin{align}
	a &= A_1^2 + A_2^2,\\
	ar = a \cos(2\alpha) &=  A_1^2  - A_2^2;\\
	\end{align}
\end{subequations}
and Eq.~\eqref{vertex_renorm} becomes
\begin{equation}
\frac{d a}{d\ln\Lambda} = -2 a^2 r,\quad
\frac{d (ar)}{d\ln\Lambda} = -2 a^2.
\end{equation}
Observe that we have
\begin{equation}
(ar)^2 = a^2- C_2^2,\quad C_2 = a_0\sqrt{ 1-r_0^2} >0.
\end{equation}
Using this constant of motion, we have 
\begin{equation}
\ddot{r} = -4C_2^2 r,\quad a = \frac{C_2}{\sqrt{1 - r^2} },
\end{equation}
and the solution reads
\begin{equation}
r = r_0 \cos(2C_2 \ln \Lambda/\Lambda_0) - \frac{C_2}{a_0}\sin(2 C_2\ln \Lambda/\Lambda_0) =  \cos(2C_2 \ln \Lambda/\Lambda_0 + \theta), \quad \tan(\theta) = \frac{C_2}{a_0 r_0};
\end{equation}
hence $max(r^2) =  1$ so the expression for $a$ is well defined. Therefore
\begin{equation}
a = \frac{C_2}{ \sin(2C_2\ln \Lambda/\Lambda_0 + \theta)} = a_0 Z_a.
\end{equation}
We finally get
\begin{equation}
\frac{d\ln{ \epsilon}}{d \ln \Lambda} = -{a},\quad \frac{d \ln {m}}{d \ln \Lambda} = {a},
\end{equation}
and therefore
\begin{equation}
{\epsilon} = {\epsilon}_0 Z_m^{1/2,}\quad 	{m} = {m}_0 Z_m^{-1/2};
\end{equation}
where
\begin{equation}
Z_m =  \frac{\cot(C_2 \ln\Lambda/\Lambda_0 + \theta/2)}{\cot(\theta/2)} \to \frac{1}{1+2a_0 \ln \Lambda/\Lambda_0}\quad \text{as}\: r_0 \to 1,
\end{equation}
that is the renormalization for the scalar disorder considered in Ref.~\citen{Ado}.
If we assume $a_0 \ln\Lambda_0 \ll 1$, these expressions can be approximately written as
\begin{equation}
Z_m \approx [1 - 2 \cos(2\gamma) a_0 \ln \Lambda_0/\Lambda]^{-1}, \quad Z_a \approx [1-2a_0r_0 \log \Lambda_0/\Lambda]^{-1},
\end{equation}
and the RG process should terminate when $\Lambda^2 = ({\epsilon}^2 - {m}^2)$. Therefore both $Z_a$ and $Z_m$ are of order $1$.	 For example, for the bare parameters $E_0 = 1, m_0 = 0.8, \Lambda_0 = 1000, a_0 = 0.001, r_0 =0.5, s_0= 0.2 $, we get $Z_m = 1.0147, Z_a = 1.0075$ and the renormalized values for relative strength and covariance becomes $s = 0.2015, r=0.5108$.

\subsection{Once corrected current vertices}
\label{sec:A}
Furthermore, the disorder lines introduce vertex corrections as in Fig~\ref{fig:self-energy-vertex}. 
If we define the disorder vertex operators
\begin{subequations}
	\label{vertex_functions}
	\begin{align}
	\label{scalar-gauge}
	\Gamma_+ &= \cos(\gamma) + \sin(\gamma)\sigma_x,\\
	\Gamma_- &= \cos(\alpha) \sigma_z + \sin(\alpha)\sigma_x.
	\end{align}  
\end{subequations}
The once corrected current operators are
\begin{equation}
\label{once_corrected}
\sigma_\mu^{\pm} = \frac{1\pm r}{\tau}  \int \frac{d^2 k}{(2\pi)^2} \Gamma_\pm G^R \sigma_\nu G^A \Gamma_\pm  = \frac{1\pm r}{2}\left\langle \frac{ \Gamma_\pm g^R \sigma_\nu g^A \Gamma_\pm }{2|\epsilon^2 + b m^2 + ck_x|}\right\rangle_{FS} = \frac{1 \pm r}{2}(A^{(\pm)})^\nu_\mu \sigma_\nu.
\end{equation}
Here, the Greek indices run from $\mu = 1...4$, where $\sigma_{1,2,3} = \sigma_{x,y,z}$ and $\sigma_4 = \mathbf{1}_{2\times 2}$. The small case letter $g$ is the numerator of the Green's function and the symbol $\langle...\rangle_{FS}$ means Fermi surface averaging, i.e. setting the magnitude of momentum to $k_F = \sqrt{\epsilon^2 -m^2}$ and performing an angular averaging. 	We use the convention, where the upper index refers to the row number while the lower index refers to columnn number. By defintion, the tensors $A^{(\pm)}$ transforms the bare vertex to the vertex once corrected by disorder vertices $\Gamma_\pm$ respectively. They read
\begin{equation}
(A^{(\pm)})^\nu_\mu  =  \frac{1}{2}\text{tr}\left [ \sigma_{\nu} \left\langle \frac{ \Gamma_\pm g^R \sigma_{\mu} g^A \Gamma_\pm }{2| \epsilon^2 + b m^2 + ck_x|}\right\rangle\right]
= \left\langle\frac{R^{(s)\nu}_\mu}{ |\epsilon^2 + b m^2 + c k_x|}\right\rangle = \frac{1}{\Omega} \left\langle\frac{R^{(s)\nu}_\mu}{1 + \cos(\theta) \cos(\phi)}\right\rangle_{FS}, 
\end{equation}
where $R$ is a tensor defined here just to encapsulate the numerator and streamline the equations and
\begin{equation}
\Omega =  \epsilon^2 + b m^2 > 0 ,\quad \cos(\theta) = \frac{c k_F}{ \epsilon^2 + b m^2}.
\end{equation}
and $FS$ denotes Fermi surface average. 
The Fermi surface average requries the following integrals
\begin{equation}
\frac{1}{2\pi} \int_0^{2\pi} \frac{d\phi (1,\cos(\phi),\cos(2\phi))}{1+\cos(\theta)\cos(\phi)},
\end{equation}
the first one of which evaluates to $\csc(\theta)$ and the others can be easily computed from this result.
Notice that if we transpose $A^{(s)}$, this amounts to interchanging $g^R$ and $g^A$ in the trace and introduce a minus sign whenever $\sigma_{\mu}$ or $\sigma_\nu$ is  equal to $\sigma_y$. However, $g^R- g^A \sim 1/\tau$, therefore we must have $A$  of the form
\begin{equation}
T A^{(\pm)} T =
\begin{pmatrix}
A^{(\pm)}_{11} & 0 \\ 
0 & \bar{A}^{(\pm)}_{3\times 3}
\end{pmatrix}
+ \frac{1}{\tau} 
\begin{pmatrix}
0 & v_1^{(\pm) T} \\ v_2^{\pm} & 0_{3\times 3}
\end{pmatrix},
\end{equation}
where 
\begin{equation}
T=
\begin{pmatrix}
\sigma_x & 0 \\ 0 & 1_{2\times 2}
\end{pmatrix},
\end{equation}
which helps extract results at given order in $\tau$, at various stages. 

In addition to the transformation tensors $A^{\pm}$ due to each component of disorder $\Gamma_\pm$, we have the total transformation tensor $A$, that is the weighted sum $A = (1+r) A^{+} + (1-r) A^{-}$. 

The transformation tensors can be computed explicitly to yield simple form when the disorder potential assumes the limiting values in Table~\ref{table:limits} of the main text. When we have uncorrelated charge and mass disorder only, i.e. $\alpha = \gamma = 0$, $A$ separates into $2\times 2$ blocks and the relevant block that corrects $\sigma_{x}$ and $\sigma_y$ reads
\begin{equation}
\label{relevant_block}
A_{2\times 2}^{(\pm)} = \frac{ 1}{\epsilon^2+m^2}\begin{pmatrix}
\frac{\epsilon^2-m^2}{2} & -\frac{\epsilon m}{\tau} \\ 
\frac{\epsilon m}{\tau} & \frac{\epsilon^2-m^2}{2}.
\end{pmatrix}
\end{equation}
$A^{(+)}$ is what we get if we only had scalar disorder.\cite{Ado}

When $\hat{V} = \sigma_x$, one can go to the limit $r = 1$ and $\gamma = \pi/2$, which produces $A_{2\times 2} = \text{diag}(1/2,-1/2)$. 

When $\hat{V} = \mathbf{1} + \sigma_x$, we go to the limit $r= 1$ and $\gamma = \pi/4$, and get
\begin{equation}
A =  \begin{pmatrix}
0 & 0 & m & 1\\
0 & 0 & 0 & 0\\
0 & 0 & 0 & 0\\
0 & 0 & m & 1	
\end{pmatrix}.
\end{equation}

Finally, when $\hat{V} = \mathbf{1}  +\sigma_z $, we use the transformation at the Lagrangian level described in the paper between  in Sec.~\ref{app:lagrangian_absorption} of this supplement, at the value $\beta=\pi/4$ and work with the transformed values $\tilde{m}=\tilde{\epsilon}$ and $\tau \to \infty$. This immediately yields $A_{2\times 2} = 0$. This can also be seen by explicitly calculate the vertex correction as in Eq.~\eqref{once_corrected}. Note that the term $\Gamma g^R \sigma_x g^A \Gamma$ vanishes for $\Gamma = \mathbf{1} + \sigma_z$, simply because once can move one of the $\Gamma$'s all the way to the other side to get $(\mathbf{1} -\sigma_z)(\mathbf{1}+\sigma_z) =0$.

\subsection{Full corrected vertex}
Writing $A = \frac{1+r}{2} A^{(+)} + \frac{1-r}{2} A^{(-)}  $,
the full corrected vertex $\bar{\sigma}_\mu = \mathcal{A}^\nu_\mu \sigma_\nu$ can be found by solving
\begin{equation}
\left( \delta_\mu^\nu -  A^{\nu}_\mu\right) \mathcal{A}_{\alpha}^\mu   = \delta^\nu_\alpha .
\end{equation}
However, the matrix $\mathbf{1} - A$ is singular  (due to the fact that the spin density is always Dirac mass times charge density). If we define
\begin{equation}
C = \mathbf{1}-A,
\end{equation}
the solutions to the indeterminate system are
\begin{subequations}
	\begin{align}
	\mathcal{A}_{1}^\nu &= (C^{\pmb{+}})_1^\nu + [1-C^{\pmb{+}}C]\xi,\\
	\mathcal{A}_{2}^\nu &= (C^{\pmb{+}})_2^\nu + [1-C^{\pmb{+}}C]\xi,
	\end{align}
\end{subequations}
where $\xi$ is an arbitrary matrix and $C^{\pmb{+}}$ (bold \pmb{+} sign not to be confused with + that denotes the scalar-gauge component of disorder as in Eq.~\eqref{scalar-gauge}) is the Moore-Penrose pseudo-inverse of $C$. There is no solution for the other vertices $\bar{\sigma}_3,\bar{\sigma}_4$, however any observable that contain these vertices describe response due to a constant electric potential and must vanish due to gauge invariance.  
We should also check that while calculating any
observable, the arbitrary component does not contribute, which we do for conductivity. 

For the simpler situation where $\alpha = \gamma = 0$, $A$ separates into blocks and the relevant part Eq.~\eqref{relevant_block} is invertible. The full corrected vertices $\bar{\sigma}_3,\bar{\sigma}_4$ can still not be computed as their relevant block is singular, a consequence of gauge invariance.


\subsection{Non-crossed diagram}
The non-crossed diagram (see the top left corner of Fig.~\ref{fig:bubbles}) contribution to Hall conductivity is
\begin{equation}
{\sigma}^{nc}_{xy} = \frac{1}{2\pi}\text{tr}\int \frac{d^2 k}{(2\pi)^2} \sigma_x G^R \bar{\sigma}_y G^A \\= \frac{1}{2\pi}\mathcal{A}^\mu_2  \frac{\tau}{2} \text{tr}\sigma_x \frac{2}{\tau}\int \frac{d^2 k }{(2\pi)^2} G^R \sigma_\mu G^A  \\=\frac{1}{2\pi} \mathcal{A} ^\mu_2  \frac{\tau}{2} \text{tr}\sigma_x ( A^{(0) } )^\nu_\mu \sigma_\nu,
\end{equation}
where $A^{(0)}$ is the vertex function due to the scalar disorder vertex, but all disorder contribution is included in self-energy.
Therefore we have
\begin{equation}
\sigma_{xy}^{nc} =   \frac{\tau}{2\pi} (A^{(0)}\mathcal{A})_2^1.	
\end{equation}
This is a fairly simple numerical computation and the arbitrary solution can be checked to have no contribution in other words, $(A^{(0)}[1-C^{\pmb{+}}C])^1_\nu$ is a zero vector up to working order.

Note that moving the ladder from one vertex to the other does not change the result. This is captured by 
\begin{equation}
\text{tr}[\sigma_x G^R \bar{\sigma}_y G^A ] = 	\text{tr}[(\bar{\sigma}_x)^* G^R \sigma_y G^A ].
\end{equation}
Here in $({\sigma}_x)^*$, the conjugation interchanges  the internal $G^R,  G^A$ lines. (If we were to compute the  left corrected  $\sigma_y$  an additional minus sign should be inserted due to conjugation).
Because this expression has to be real, we can simply conjugate it to get
\begin{equation}
\sigma_{xy} = \frac{1}{2\pi} \text{tr}[(\bar{\sigma}_x)^* G^R \sigma_y G^A ]  =  -\frac{1}{2\pi} \text{tr}[\bar{\sigma}_x G^A \sigma_y G^R ] = -\sigma_{yx}.
\end{equation}
An identity that holds for all diagrams	, irrespective of whether disorder breaks rotational symmetry.

Finally for longitudinal conductivities we have
\begin{subequations}
	\begin{align}
	\sigma_{xx} &= \frac{\tau}{2\pi} (A^{(0)}\mathcal{A})_1^1,\\
	\sigma_{yy} &= \frac{\tau}{2\pi} (A^{(0)}\mathcal{A})_2^2.
	\end{align}
\end{subequations} 

Theses conductivities computed explicitly to yield simple form when the disorder potential assumes the limiting values in Table~\ref{table:limits} of the main text. When we have uncorrelated charge and mass disorder only, i.e. $\alpha = \gamma = 0$, we simply have $A^{(0)} = A^{(+)}$. Then the contribution to transverse conductivity follows as
\begin{equation}
\sigma_{xy}^{nc.} = -\frac{2}{\pi} \frac{\epsilon m (\epsilon^2 +m^2)}{([2-r]\epsilon^2 + [2+r]m^2)^2}.
\end{equation} 
The non-crossed contribution captures everything when it comes to longitudinal conductivity (in this case isotropic), where the effective parameters are same as the originals, as there are no correlations.

When $\hat{V} = \sigma_x$, one can go to the limit $r = 1$ and $\gamma = \pi/2$, which produces $A^{(0)}_{2\times 2} = \text{diag}(1/2,1/2)$. We know from Sec.~\eqref{sec:A}, $\mathcal{A}_{2\times2} = (\mathbf{1}-A_{2\times 2})^{-1} = \text{diag}(2, 2/3)$. Since both $A^(0)$ and $\mathcal{A}$ are diagonal, the transverse conductivity vanishes. On the other hand, the longitudinal conductivities are $\sigma_{xx} = \tau/(2\pi), \sigma_{yy} = \tau/(6\pi)$. The unisotropy stems from the fact that the gauge field disorder picks a direction.

When $\hat{V} = \mathbf{1} + \sigma_x$, we go to the limit $r= 1$ and $\gamma = \pi/4$, and get
\begin{equation}
(A^{(0)}\mathcal{A})_{2\times 2} =
\begin{pmatrix}
(\epsilon-m)/m &-1/(2\tau)\\
1/(2\tau) & (\epsilon-m)/\epsilon
\end{pmatrix}.
\end{equation}
Hence, the contribution to transverse conductivity assumes its universal value $-1/(4\pi)$.

Finally, when $\hat{V} = \mathbf{1}  +\sigma_z $, we use the transformation in Sec.~\ref{app:lagrangian_absorption} of this supplement,  where $\beta = \pi/4$ and $r =  \tilde{r} = 0$ and work with the transformed values $\tilde{m}=\tilde{\epsilon}$ and $\tilde{\tau} \to \infty$. Calculating the subtle limit we find out that the contribution to transverse conductivity assumes its universal value $-1/(4\pi)$ and the isotropic transverse conductivity is $\sigma_{xx} = \tau/(4\pi) \times (\epsilon-m)/(\epsilon+m)$. 

Having computed the non-crossed diagram contribution to conductivities, we will now focus on crossed diagrams. 

\subsection{Crossed diagrams}
\label{sec:crossed}
It is advantageous to calculate the crossed-diagrams (see the diagrams with crossing dashed lines in Fig.~\ref{fig:bubbles}) in momentum space rather than real space as done in Ref.~\citen{Ado}. We do this by using symmetry behaviour of the diagrams under complex conjugation and interchange of vertices.
\subsubsection{X-diagram}
The X diagram (top right corner in Fig:\ref{fig:bubbles}) in momentum space reads
\begin{equation}
\sigma_{xy}^X = \frac{g^2}{2\pi} \sum_{a,b} w^a w^b \int \frac{d^2 \mathbf{k}}{(2\pi)^2} \frac{d^2 \mathbf{Q}}{(2\pi)^2} \frac{d^2 \mathbf{q}}{(2\pi)^2} 
\\ \text{tr} [\bar{\sigma}_x G^R(\mathbf{k})\Gamma_a G^R(\mathbf{Q}+\mathbf{q})\Gamma_b G^R(\mathbf{Q})\bar{\sigma}_y G^A(\mathbf{Q})\Gamma_a\\ G^A(\mathbf{k}-\mathbf{q})\Gamma_b G^A(\mathbf{k}) ],
\end{equation}
where $a,b = +,-$ and $w = (1+r,1-r)/2$ are the weights of vertices in Eq.~\eqref{vertex_functions}. 
The Hall conductivity must be real and this can easily seen by conjugate transposing the expression inside trace, which produces the same integrand after a relabeling $a \leftrightarrow b$ and $\mathbf{q} \to \mathbf{k -Q -q}$.

The vertices $G^A(\mathbf{k}) \bar{\sigma}_x G^R(\mathbf{k})$ and $G^R(\mathbf{Q}) \bar{\sigma}_y G^A(\mathbf{Q})$ allows us to integrate over the magnitudes  $k,Q$ and fixes them to $k_F = \sqrt{\epsilon^2-m^2}$. After this operation the transverse conductivity becomes
\begin{multline}
\frac{h}{e^2}\sigma_{xy}^{X}=\frac{1}{4\Omega^2} \sum_{a,b} w^a w^b 	\int  \frac{d \phi_k}{2\pi} \frac{d \phi_Q}{2\pi} \frac{d^2 \mathbf{q}}{(2\pi)^2} \text{tr}\left[\bar{\sigma}_x g(\hat{k} k_F)\Gamma_a g(\hat{Q}k_F+\mathbf{q}) \Gamma_b g(\hat{Q}k_F)\bar{\sigma}_y g(\hat{Q}k_F)\Gamma_a g(\hat{k}k_F-\mathbf{q})\Gamma_b g(\hat{k}k_F)\right]\times\\
\frac{1}{1+\cos(\theta)\cos(\phi_k)}\frac{1}{1+\cos(\theta)\cos(\phi_Q)}\times
\frac{1}{q^2 - 2qk_F \cos({\phi}_k-\phi_q) +i/\tau}	\frac{1}{q^2 + 2qk_F \cos({\phi}_Q-\phi_q ) -i/\tau }.
\end{multline}
Let us call the numerator and denominator of the integrand $F$ and $P$ respectively. Multiplying all angles by $-1$ and conjugating the integrand switches the sign of the numerator and conjugates the denominator.This means only 
\begin{equation}
-  \text{Im} F \text{Im}P^{-1} = - f \text{Im} P^{-1}
\end{equation}
contributes.
Furthermore, a simple relabeling $\phi_Q \leftrightarrow \phi_k$ and $\phi_q \to \phi_q+\pi$ results in $\bar{\sigma}_x \leftrightarrow \bar{\sigma}_y$  and $P \to P^*$ . This means the result depends only on the part of  $f$ that is  antisymmetric over the interchange $\bar{\sigma}_x \leftrightarrow \bar{\sigma}_y$.	
Meanwhile  the part of the trace that does not depend on $q$, after anti-symmetrizing over $\bar{\sigma}_x \leftrightarrow \bar{\sigma}_y$,   is antisymmetric under $\phi_Q \leftrightarrow \phi_k$. However, in the limit $q \to 0$, the denominator is symmetric under $\phi_Q \leftrightarrow \phi_k$, therefore this component must vanish. Furthermore, the term quadratic in $q\cdot \sigma$ when conjugate transposed under trace and upon the interchange of $a\leftrightarrow b$ is invariant,  however because this trace changes sign after conjugation, this term must also vanish.

All in all, if we define
\begin{equation}
J_\mu^{ab}(k) = \Gamma_a g(k) \bar{\sigma}_\mu  g(k) \Gamma_b,
\end{equation} 
we can compute the necessary imaginary part only 
\begin{equation}
f =  \sum_{ab}w^a w^b \text{Im} \text{tr}[\hat{q}\cdot\pmb{\sigma} J_y^{ba}(\mathbf{Q}) g(\mathbf{k})  J_x^{ba}(\mathbf{k}) ] + (\mathbf{Q} \leftrightarrow \mathbf{k} );
\end{equation}
and all other terms can be generated through the above 	 symmetry operations. One can carry out the $q$ integral rather easily, because the imaginary part of $P^{-1}$ is made of delta functions of $q$. After this we get		
\begin{equation}
\frac{h}{e^2}\sigma_{xy}^{X}= \frac{1}{16 \Omega^2 k_F} PV \int  \frac{d \phi_k}{2\pi} \frac{d \phi_Q}{2\pi} \frac{d \phi_q}{2\pi}
\frac{1}{1+\cos(\theta)\cos(\phi_k)}\\ \times \frac{1}{1+\cos(\theta)\cos(\phi_Q)}
\frac{f}{\cos(\phi_k-\phi_q) +\cos(\phi_Q-\phi_q)}. 
\end{equation}
Here, the principal value of the integral is computed on the domain, from which the points whereever $\phi_k = \phi_Q \pm \pi$, at which the denominator becomes real and the integral must vanish.   
The integral over $\phi_q$ is trivial after computing $f$ and noticing that, due to the factor $\hat{q}\cdot \sigma$ and the symmetry under $Q\leftrightarrow k $ it can be broken down into
\begin{equation}
f = 2\cos(\phi_q)\cos(\phi_+) \tilde{f} + 2 \sin(\phi_q) \sin(\phi_+) \tilde{f},
\end{equation}
where $\phi_Q = \phi_+ + \phi_-$ and $\phi_k = \phi_+ - \phi_-$.
Finally we get
\begin{equation}
\frac{h}{e^2}\sigma_{xy}^{X}= \frac{1}{16 \Omega^2 k_F} PV \int  \frac{d \phi_k}{2\pi} \frac{d \phi_Q}{2\pi} \frac{\tilde{f}}{ \cos \left(\frac{\phi_Q-\phi_k}{2} \right)} \\ \times \frac{1}{1+\cos(\theta)\cos(\phi_k)}\frac{1}{1+\cos(\theta)\cos(\phi_Q)}.
\end{equation}
For the scalar case, the unisotropy angle $\theta = \pi/2$, and
\begin{equation}
\tilde{f}^{scalar.dis.} = \frac{4 \Omega^2}{(\epsilon^2 +3m^2)^2}\times 32 m k_F^3 \cos(\phi_-) \sin^2(\phi_-)
\end{equation}
(the coefficients come from the disordered corrected vertices $\bar{\sigma}_x,\bar{\sigma}_y$). Hence the result follows as
\begin{equation}
\sigma_{xy}^{scalar.dis.} = \frac{e^2}{h}\frac{4 m k_F^2}{(\epsilon^2+3m^2)^2}.
\end{equation}

For the general case, the principal value must be computed.
This can be done by subtracting the term that is non-integrable on the whole domain but has zero principal value:
\begin{equation}
\frac{1}{32 \Omega^2 k_F} PV \int  \frac{d \phi_k}{2\pi} \frac{d \phi_Q}{2\pi} \frac{\tilde{f}(\phi_- = \pi/2)}{ \cos(\phi_-)}  \frac{1}{1-\cos^2(\theta)\sin^2(\phi_+)}=0.
\end{equation}
Up on subtracting this  term the integral for the $X$ diagram can be computed on the whole integration domain through a straight forward numerical procedure. 

\subsubsection{$\Psi$-diagrams}
Similarly, the $\Psi$ diagrams (bottom two diagrams in Fig:\ref{fig:bubbles}) are written as
\begin{equation}
\sigma_{xy}^\Psi = \frac{g^2}{2\pi} \sum_{a,b} w^a w^b\int \frac{d^2 \mathbf{k}}{(2\pi)^2} \frac{d^2 \mathbf{Q}}{(2\pi)^2} \frac{d^2 \mathbf{q}}{(2\pi)^2}  \text{tr} [\bar{\sigma}_x G^R(\mathbf{k})\Gamma_a G^R(\mathbf{k-q})\Gamma_b G^R(\mathbf{Q-q})\Gamma_a 
\\ G^R(\mathbf{Q})\bar{\sigma}_y G^A(\mathbf{Q})\Gamma_b G^A(\mathbf{k})] + h.c.,
\end{equation}
an expression that is necessarily real because its hermitian conjugate is added to itself. 

This can be decomposed as
\begin{multline}
\frac{h}{e^2} \sigma_{xy}^{\Psi} = \frac{1}{4 \Omega^2}\sum_{ab} w^a w^b \int   \frac{d\phi_k}{2\pi}\frac{d\phi_Q}{2\pi} \frac{d^2 \mathbf{q}}{(2\pi)^2} \text{tr}\left[\bar{\sigma}_x g(\hat{k}k_F)\Gamma_a g(\hat{k}k_F-\mathbf{q}) \Gamma_b g(\hat{Q}k_F-\mathbf{q}) \Gamma_a g(\hat{Q}k_F)\bar{\sigma}_y g(\hat{Q}k_F) \Gamma_b g(\hat{k}k_F)\right]\\\times  \frac{1}{1+\cos(\theta)\cos(\phi_k)}\frac{1}{1+\cos(\theta)\cos(\phi_Q)}\times
\frac{1}{q^2 - 2qk_F \cos({\phi}_k-\phi_q) -i/\tau} 	\frac{1}{q^2 - 2qk_F \cos({\phi}_Q-\phi_q ) -i/\tau } + h.c.
\end{multline}
Multiplying all angles by $-1$  and conjugating the integrand switches the sign of the numerator $F = \text{Re} F + i f$  and conjugates the denominator $P$ . However trace of hermitian conjugate is equal to the trace of complex conjugate, therefore the sum of two complementary diagrams should yield $-2  f \text{Im}P^{-1}$.

Interchanging $\bar{\sigma}_x \leftrightarrow \bar{\sigma}_y$  simply leads to $f \leftrightarrow f^*$, upto $k \leftrightarrow Q $, hence produces   a minus sign.

Note also, that the parts of the integrand that are $\sim q^0$ must be dropped, because for $q=0$, the denominator becomes real. So we can write $f = \tilde{f} q + \tilde{f}_2 q^2$.

When $\cos(\phi_k - \phi_q) = \cos(\phi_Q - \phi_q)$, the integrand has a double pole, and the $q$ integral can be written as
\begin{equation}
\text{Im}\int_0^\infty \frac{dq}{2\pi} \frac{q f}{[q^2-2qk_F \cos(\phi_k - \phi_q)-i/\tau]^2 } = \pi \int_0^\infty \frac{dq}{2\pi} \frac{1}{q}\partial_q \left(\frac{q f}{2q - 2k_F\cos(\phi_k-\phi_q)}\right) \delta(q- 2k_F\cos(\phi_k-\phi_q)) = \frac{1}{2} \tilde{f}_2|.
\end{equation}
If the double pole is caused by $\phi_k = \phi_Q$, this quantity vanishes up on integrating over $\phi_q$. If it is due to $\phi_q = \phi_+ = (\phi_Q + \phi_k)/2$, we should integrate it with respect to the angles to find the contribution, however this contribution will be taken into account automatically because there is a pole-zero cancellation, as we shall see below.

The integral on the rest of the domain can be calculated as follows. Writing the denomiators as 
\begin{equation}
\frac{1}{q^2 -2 q k_F\cos(\phi_k)} -i\pi\frac{1}{2k_F|\cos(\phi_k)|}
\times[ \delta(q-2k_F \cos(\phi_k)) + \delta(q) ] + (Q\leftrightarrow k),
\end{equation}
we have 
\begin{equation}
\frac{h}{e^2} \sigma_{xy}^{\Psi} =-\frac{1}{4\Omega^2 k_F}PV\int  \frac{d\phi_k}{2\pi}\frac{d\phi_Q}{2\pi}\frac{d\phi_q}{2\pi}dq \frac{[\tilde{f} + \tilde{f}_2 q]\delta(q-2k_F\cos(\phi_k - \phi_q))}{\cos(\phi_k-\phi_q)- \cos(\phi_Q-\phi_q)} \times \frac{1}{1+\cos(\theta)\cos(\phi_k)}\frac{1}{1+\cos(\theta)\cos(\phi_Q)}
\end{equation}
where $f$ is already assumed to be symmetrized with respect to $Q\leftrightarrow k$. 
After the $q$ integral what remains from the delta function is the constraint that $\cos(\phi_k - \phi_q)>0$,  however	 with $\phi_q \to \phi_q + \pi$, both the numerator and the denominator switches sign, while the constraint becomes $\cos(\phi_k - \phi_q)<0$ hence this constraint disappears. 
\begin{equation}
\frac{h}{e^2} \sigma_{xy}^{\Psi} =-\frac{1}{8\Omega^2 k_F}PV\int  \frac{d\phi_k}{2\pi}\frac{d\phi_Q}{2\pi}\frac{d\phi_q}{2\pi} \frac{\tilde{f} + 2k_F \cos(\phi_k-\phi_q) \tilde{f}_2 }{\cos(\phi_k-\phi_q)- \cos(\phi_Q-\phi_q)} \times \frac{1}{1+\cos(\theta)\cos(\phi_k)}\frac{1}{1+\cos(\theta)\cos(\phi_Q)}.
\end{equation}  
Symmetrizing over $\phi_Q \leftrightarrow \phi_k$ we get 
\begin{equation}
\frac{h}{e^2} \sigma_{xy}^{\Psi} =-\frac{1}{4\Omega^2 }\int  \frac{d\phi_k}{2\pi}\frac{d\phi_Q}{2\pi}\frac{d\phi_q}{2\pi}  \tilde{f}_2 \times \frac{1}{1+\cos(\theta)\cos(\phi_k)}\frac{1}{1+\cos(\theta)\cos(\phi_Q)},
\end{equation}
where
\begin{equation}
\tilde{f}_2 = \sum_{ab} w^a w^b \text{Im}\:\text{tr}[\bar{\sigma}_x g(\hat{k}k_F)\Gamma_a \hat{q}\cdot \sigma \Gamma_b \hat{q}\cdot \sigma \Gamma_a g(\hat{Q}k_F)\\\bar{\sigma}_y g(\hat{Q}k_F) \Gamma_b g(\hat{k}k_F)].
\end{equation}
For the scalar disorder case, this function vanishes, so does $\sigma^\Psi$. The $\phi_q$ integral can be carried out easily, as the only place this variable appears is in $\hat{q}\cdot \pmb{\sigma}$. 	The rest of the integral can be computed by using a straight forward numerical procedure. 

The above calculations in Sec.~\ref{sec:self-energy}-\ref{sec:crossed} allows us to start from the disorder model in Eq.~\eqref{disorder_transformed} and obtain the contribution of each diagram to the total transport coefficient. 
\begin{figure}[H]
	\centering
	\begin{subfigure}{0.3\textwidth}
		\centering
		\includegraphics[width=0.6\textwidth]{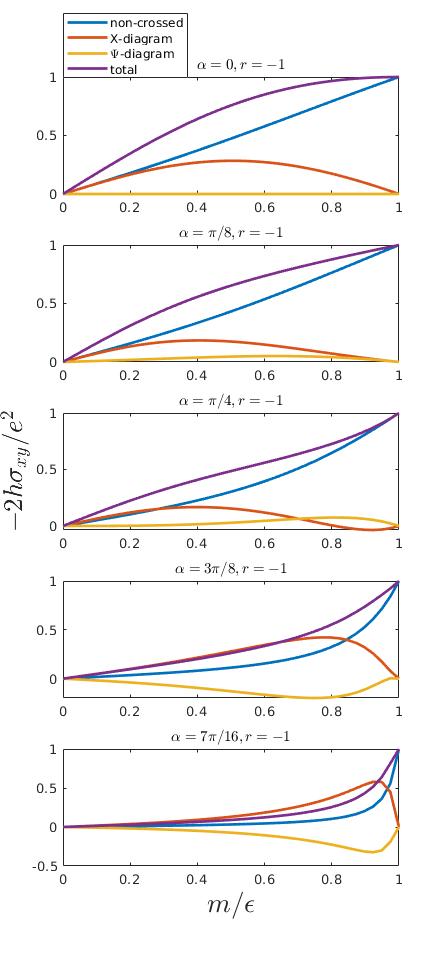}
		\caption{ }\label{fig:alpha}
	\end{subfigure}
	\begin{subfigure}{0.3\textwidth}
		\centering
		\includegraphics[width=0.6\textwidth]{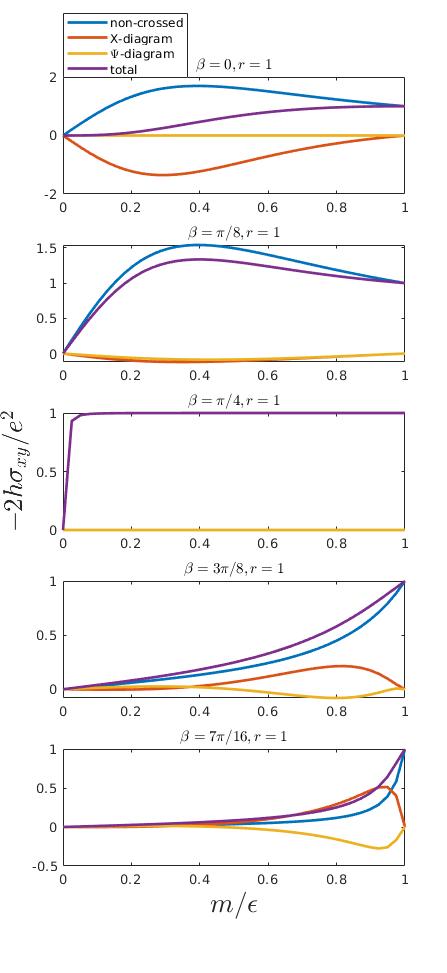}
		\caption{}\label{fig:beta}
	\end{subfigure}
	\caption{a) Transverse conductivity (sign inverted and measured in $e^2/(2h)$)as a function of Dirac mass for different values of effective tilting angle $\alpha$, measured from surface normal. The disorder potential is given as Eq.~\eqref{disorder_transformed}. The scalar component of disorder that is unrelated to magnetic impurities is assumed zero.b)Transverse conductivity as a function of Dirac mass for a different values of effective correlation angle $\gamma$. The disorder potential is given as Eq.~\eqref{disorder_transformed}. The scalar component of disorder that is unrelated to magnetic impurities is assumed very strong, yet the scalar disorder is highly correlated with the gauge field disorder. Refer to the legend in the previous page.}
\end{figure}
For example, Fig.~\ref{fig:alpha}, we see the diagrams of $\sigma_{xy}$ and the total AHE as the effective tilting angle is varied for a correlated disorder with mass and gauge disorder only. We observe tha the contribution of the crossed diagrams is small and there is a general suppression of conductivity as the effective tilting angle grows towards in plane. 

However, the effect of a growing effective correlation angle for scalar and gauge correlated disorder is more drammatic as seen in Fig.~\ref{fig:beta}. As soon as the effective correlation angle grows above zero, the counter acting effect of crossed diagrams to skew-scatterig dies out and this leads to an increase in Hall conductivity. As discussed by previous authors, the non-crossed contribution exceeds the universal value of QAHE for a metallic system, however, in scalar case $(\gamma)$ AHE is suppressed by the skew-scattering contribution. Moderate values of effective correlation destroys this suppression up until $\gamma = \pi/4$ where the AHE is purely intrinsic and equal to the universal result. Further  increase in the correlation makes the disorder potential closer to pure gauge, disorder hence AHE is suppressed when $\gamma$ exceeds $\pi/4$ and approaches $\pi/2$. 

\section{`Sign problem' in the AHE of magnetically doped topological insulator thin films}

\begin{table}[ht]
	\begin{tabular}{llllll}
		Experiment                                                   & Dopant(\%) & Compound       & $\text{sgn}(\sigma_{xy})$ & Phase     & Carrier type  \\ 
		\hline
		Ref.~\citen{Checkelsky2012} & Mn (\%4)                 & \ce{Bi2(Te,Se)3}    &   -    & Metal     & N-type        \\
		Ref.~\citen{LeePRB2014}          & Mn(\%2-10)               & \ce{Bi2Te3}         &   -    & Metal     & N-type        \\
		Ref.~\citen{Liu2018}             & Mn (\%1-9)                 & \ce{Bi2Se3}  &   -    & Metal     & N-type        \\
		Ref.~\citen{Zhang1582,ChangPRL2014}      & Cr (\%4-22)                & \ce{Bi2Se3}         &   -    & Metal     & N-type        \\
		Ref.~\citen{Zhang1582}           & Cr (\%22)                & \ce{Bi2Te3}         & +    & Metal     & N-type        \\
		Ref.~\citen{Zhang2014}           & Cr (\%15-22)                & \ce{(Bi_xSb_{1-x})2Te3}    & +    & Metal     & N$\rightarrow$ P          \\
		Ref.~\citen{QAHE_exp2}          & V(\%13)                  & \ce{Sb2Te3}         & +    & Insulator & NA            \\
		Ref.~\citen{QAHE_exp1,QAHE_exp3}          & Cr(\%12)                 & \ce{(Bi,Sb)2Te3}    & +    & Insulator & NA            \\
		Ref.~\citen{experimental_AHE}   & Cr(\%15)                 & \ce{(Bi,Sb)2Te3}    & +    & Insulator & NA           
	\end{tabular}
	\caption{Experimentally observed sign of AHE for various magnetic dopants and their concentrations.}
	\label{tabel:sign_problem}
\end{table}


\end{document}